\documentclass[journal=jacsat,manuscript=article]{achemso}

\usepackage{chemformula} % Formula subscripts using \ch{}
\usepackage[T1]{fontenc} % Use modern font encodings
\usepackage{graphicx}
\usepackage{comment}
\usepackage{hyperref}
\usepackage{algorithm}
\usepackage{siunitx}
\usepackage{algpseudocode}
\usepackage{amsmath,amssymb}
\definecolor{mygreen}{RGB}{0, 140, 0}
\definecolor{myblack}{RGB}{0, 0, 0}
\newcommand{\added}[1]{\textcolor{myblack}{#1}}
\author{Soumendu Bagchi}
\email{bagchis@ornl.gov}
\affiliation[CNMS]
{Center for Nanophase Materials Sciences, Oak Ridge National Laboratory, Oak Ridge, TN, 37831, USA}

\author{Ankita Biswas}
\affiliation[1]
{Department of Materials Science and Engineering, University of Virginia, Charlottesville, Virginia, 22902, United States}

\author{Prasanna V. Balachandran}
\affiliation[1]
{Department of Materials Science and Engineering, University of Virginia, Charlottesville, Virginia, 22902, United States}

\author{Ayana Ghosh}
\email{ghosha@iitm.ac.in}
\affiliation[2]
{\added{Department of Physics, Indian Institute of Technology, Madras, Chennai, 600036, Tamil Nadu, India}}

\author{P. Ganesh}
\email{ganeshp@ornl.gov}
\affiliation[CNMS]
{Center for Nanophase Materials Sciences, Oak Ridge National Laboratory, Oak Ridge, TN, 37831, USA}

\title{Towards "on-demand" van der Waals epitaxy with \added{adaptive ensemble sampling atomistic workflows}}

\keywords{American Chemical Society, \LaTeX}

\begin{document}
\begin{abstract}
% {\color{magenta}\textbf{Tentative leads/authors on the tasks, (sort of by order) are assigned but of course we shall discuss together as we make progress...}}\\

{Traditional approaches to achieve targeted epitaxial growth involves exploring a vast parameter space of thermodynamical and kinetic drivers (e.g., temperature, pressure, chemical potential etc). This tedious and time-consuming approach becomes particularly cumbersome to accelerate synthesis and characterization of novel materials with complex dependencies on local chemical environment, temperature and lattice-strains, specifically for nanoscale heterostructures of layered 2D materials. We combine the strength of next generation supercomputers at the extreme scale, machine learning and classical molecular dynamics simulations within an adaptive real time closed-loop virtual environment steered by Bayesian algorithms to enable asynchronous ensemble sampling of the synthesis space, and apply it to the recrystallization phenomena of amorphous transition-metal dichalcogenide (TMDC) bilayer to form stacked moiré superstructures under various growth parameters. \added{We show that such batch parallel Bayesian optimization-based online ensemble sampling frameworks for materials simulations can be promising towards achieving and accelerating on-demand epitaxy of van der Waals stacked moiré devices, paving the way towards a robust autonomous materials synthesis pipeline to enable discovery of unprecedented functionalities.}}

\end{abstract}

\section{Introduction}

Accelerating conventional synthesis efforts in van der Waals (vdW) heterostructures, is an important aspect to discover novel pathways which can harness unprecedented functionalities in 2D materials \cite{li2024review_2dgrowth}. Recent surge in efforts towards realizing autonomous synthesis \cite{szymanski2023autonomous, ament2021autonomous} and characterization \cite{liu2021lasser_cryst_cnms, roccapriore2022probing} frameworks is expediting such discoveries. However, narrowing down the optimal thermodynamic and kinetic parameter space is a nontrivial automation challenge, especially when gaining finer control over tuning these emerging functionality \textit{on demand} is the ultimate goal. For instance, vertically stacked van der Waals (vdW) heterostructures \cite{geim2013van} with optimal interlayer twists are extremely promising for manifesting tunable and novel functionalities at the nanoscale. These include emerging quantum phenomena, e.g., correlated electronic phases \cite{cao2018correlated}, unconventional superconductivity \cite{cao2018unconventional} and interlayer excitons \cite{rivera2018interlayer} etc. which can enable unprecedented optical and electronic properties in nanoscale devices. Traditionally, top-down methods such as mechanical manipulations of as-grown epitaxial layers (via release, transfer and stamping \cite{kim2016mech-rot} etc.) have been a preferred mode to produce heterostructures with tunable geometries. However, achieving fine control over interlayer orientation as well as ensuring high quality lattice and desired functionality at the wafer-scale can be challenging \cite{li2024review_2dgrowth, xu2021seeded_amorphoussub_laterials_recryst}, especially in the context of automated platforms.

To overcome the bottlenecks associated with conventional top-down approaches, alternative synthesis routes often exploit thermodynamically and kinetically driven pathways \cite{baek2023thermally,liu2021lasser_cryst_cnms} of rotational self-organization \cite{sutter2019chiral, sutter2019self} in vdW epitaxial systems \cite{liu2021lasser_cryst_cnms, gehring2012growthBi2Te2Se, kim2014principle_direct-vdW, lin2014direct}. It has been indeed demonstrated that bottom-up growth strategies can be leveraged to spontaneously evolve a range of rotationally aligned homo and heterostructures by tuning the driving forces e.g., substrate interactions \cite{sutter2019chiral, sutter2019self}, epitaxial strain \cite{wan2021strain, lu2022lattice}, electron irradiation \cite{roccapriore2022probing, maksov2019deep} and/or thermal annealing \cite{omambac2019temperature, de2022imaging} etc. Harnessing such growth drivers either individually or combinatorially can therefore pave the way towards scalable syntheses recipes for vdW layered structures with targeted twists or moiré patterns. 

 However, efficacy of such bottom-up techniques to design heterostructures with desired functionality, often hinges upon the ability to navigate the complex and intertwined parameter space of synthesizing vdW layered structures.%growth parameter space 
 \cite{jenks2020basic} %dominating growth. 
 In addition to nucleation and growth, the prominence of solid-state phase transitions \cite{li2021phase} and out-of-equilibrium pathways \cite{woods2016one, bianchini2020interplay} involving the effects of defects, precursors and metastable phases \cite{sutter2019self} only adds toward the complexity of the design space. To this end, the recent rise of high-throughput experimentation techniques \cite{kusne2014fly, hattrick2016perspective} integrated in closed-loop fashion with advanced artificial intelligence (AI)/machine learning (ML) platforms have shown to be promising in rapid and autonomous exploration \cite{szymanski2023autonomous} of vast synthesis-structure-property spaces \cite{ament2021autonomous}. 
 
 While active learning (AL) driven robotic discovery platforms \cite{chen2024navigating, ament2021autonomous, sheng2024autonomous} are in general tailored to facilitate synthesis aided by autonomous characterizations of target materials through optimal design of experiments, several challenges \cite{leeman2024challenges} still remain to achieve accelerated discovery of "new" functional materials. Many of these campaigns can benefit from deeper and %easily 
 accessible physical insights through the robust integration of theory, modeling and simulation capabilities \cite{sumpter2015bridge, ghosh2022bridging} in the autonomous experimentation (AE) loop. Moreover, it has been recently demonstrated how virtual environments \cite{kanarik2023LAM_process-opt} can serve as useful testbeds to rapid refinement and systematic benchmarking of algorithms and rigorous uncertainty quantification shielding the prohibitively expensive experimental data acquisition loops. 

 In the context of accelerating epitaxial growth of desired twisted vdW layered structures, existing self-driving thin-film synthesis platforms \cite{macleod2020self} could naturally exploit the predictive insights from atomistic simulation methods. In past, purely simulation driven investigations \cite{zhu2019controlling, bagchi2020rotational, srolovitz2016twisted, bagchi2020strain} have been able to propose novel strategies to design twisted metastable states in graphene and transition metal dichalcogenides (TMDCs) layers by means of lattice strain and thermally activated self-assembly pathways, ultimately predicting and motivating subsequent experimental efforts \cite{baek2023thermally, de2022imaging, kazmierczak2021strain}.

 % In addition to training machine learning models, Exascale computing platforms, powered by accelerator hardwares e.g., GPUs and semantic interconnects, atomistics simulations could now undergo a 400 folds increase in \cite{atchley2023frontier} accessing unprecedented throughput with length and timescales suitable for experimentally relevant applications.
 To bridge and guide autonomous nanoscale experimental (AE) platforms with insightful %materials
 molecular dynamics simulations, Bayesian active learning workflows \cite{kusne2020fly} exploiting the asynchronous throughput accessible at the Exascale can act as catalyst towards further acceleration in autonomous materials discovery. Moreover, the recent advances \cite{alexander2020exascale} in high performance computing (HPC) has pushed the boundaries of state-of-the art extreme scale computational frameworks \cite{mniszewski2021enabling} yielding unprecedented rates of floating point operations. 
 
 However, current ML and UQ-driven workflows for atomistics, especially dynamical trajectory datasets have not yet fully and efficiently leveraged the extreme scale computational resources, and are mostly limited in their current role to serve as initial motivation and/or offline knowledge-base to implicitly to kick start an AE platform. Few key challenges associated with these efforts include \textbf{(a)} orchestration and scheduling bottlenecks to handle large-volume of resource (node-hours), \textbf{(b)} navigating asynchronous on-line inference frameworks along with \textbf{(c)} on-the-fly data reduction---all three of which are paramount to towards achieving \textit{theory-experiment integrated closed-loop hpc-enabled AI workflows} incorporating predictive materials theory and simulation tools in automated experiments to make them truly autonomous. 
 
Inspired by recent experimental efforts to recrystallize MoS2 crystals from amorphous precursors \cite{wang2023recrystallization}, here we computationally sample optimal growth parameters to recrystallize (i.e. to evolve and reorganize long range order) a variety of target moiré heterostructures on-demand from a partially
disordered amorphous system.  We achieve this by demonstrating a resource-adaptive, on-the-fly decision making and parallelizable asynchronous Bayesian ensemble learning framework for dynamical atomistic simulations, we computationally sample optimal growth parameters to recrystallize (i.e. to evolve and reorganize long range order) a variety of target moiré heterostructures \textit{on-demand} from a partially disordered amorphous system. We apply active learning strategies using both conventional acquisition function (e.g. expected improvement) and direct posterior sampling (Thompson sampling) based Bayesian optimization (BO), and study the acceleration for asynchronous batch parallel acquisition in our on-the-fly ensemble molecular dynamics workflow environment. Excellent advantages over parallel random sampling as well as efficacy in identifying optimal growth space for evolving "on-demand" target structural phases are then shown.

% \clearpage 

\begin{figure*}[hbt!]
\centering
\includegraphics[height=0.7\linewidth,width=1.35\linewidth,angle=0]{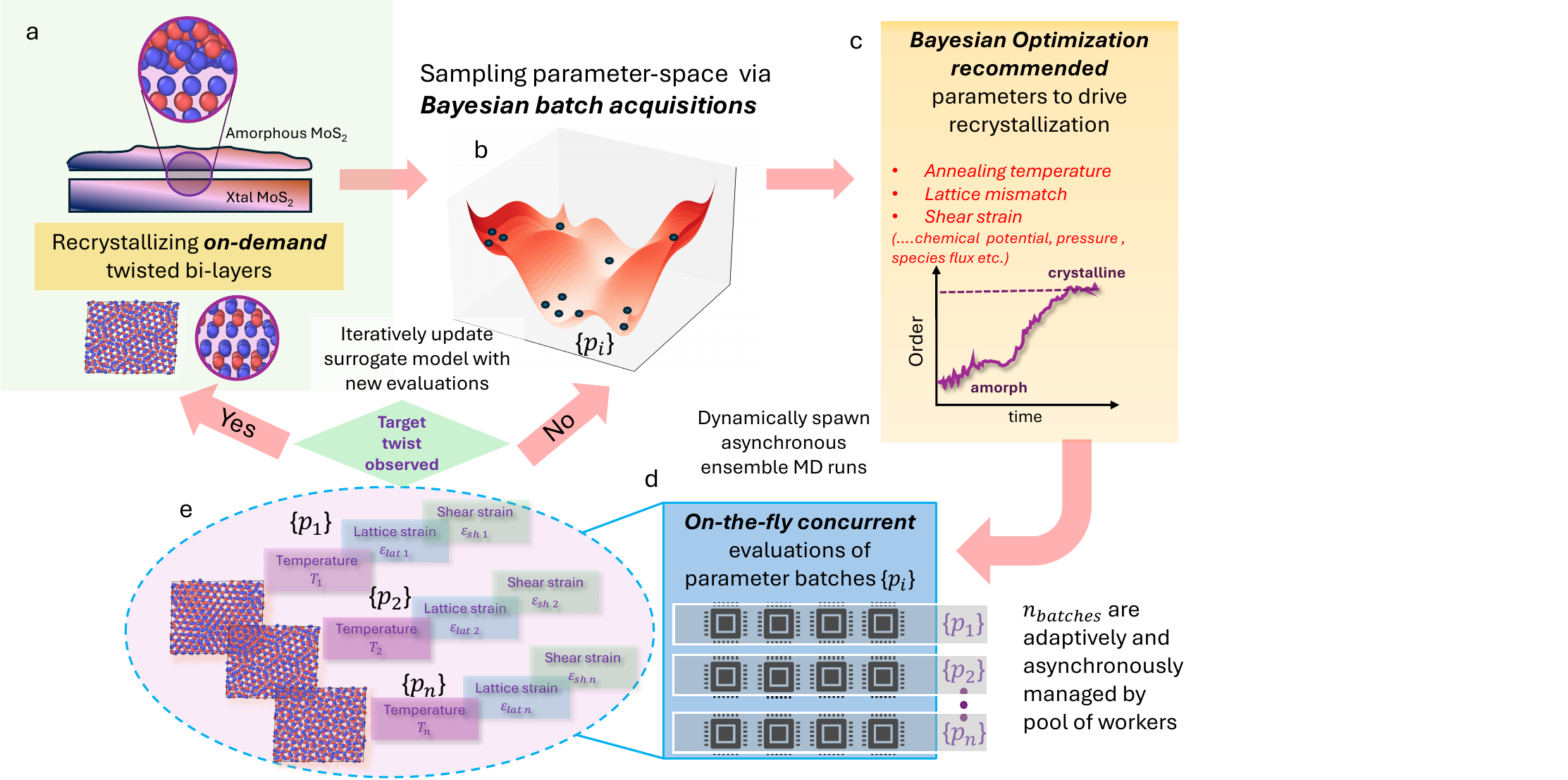}
\caption{\textbf{Asynchronous active batch sampling strategy for optimal driving  parameters for target twist angles.} Goal is to transform a given initial structure with amorphous MoS$_2$  layer on top of a crystalline layer (amorph+Xtal), into crystallized twisted bi-layer with a predefined target \textbf{(a)}. Asynchronous ensemble optimization \textbf{(b)} strategies using Bayesian acquisition of candidate recrystallization parameters \textbf{(c)}. \textit{On the fly} concurrent evaluations of atomistic simulation via a resource-adaptive dynamic orchestration under a parameter space consisting of annealing temperature, lattice and shear strains. Additionally, this space can easily be expanded (as indicated in smaller fonts) to consider other synthesis relevant parameters like chemical potential, pressure, ion/species flux etc. can naturally be \textbf{(d)} to identify the driving parameters ($\{p_i\}$) to achieve recrystallization with the target twist angle \textbf{(e)}.}
\label{fig:workflow_schem}
\end{figure*}
\section{Results}
\subsection{A real-time hpc-driven asynchronous and autonomous workflow for accelerated computational synthesis}
\begin{figure*}[hbt!]
\centering
\includegraphics[height=0.6\linewidth,width=1.\linewidth,angle=0]{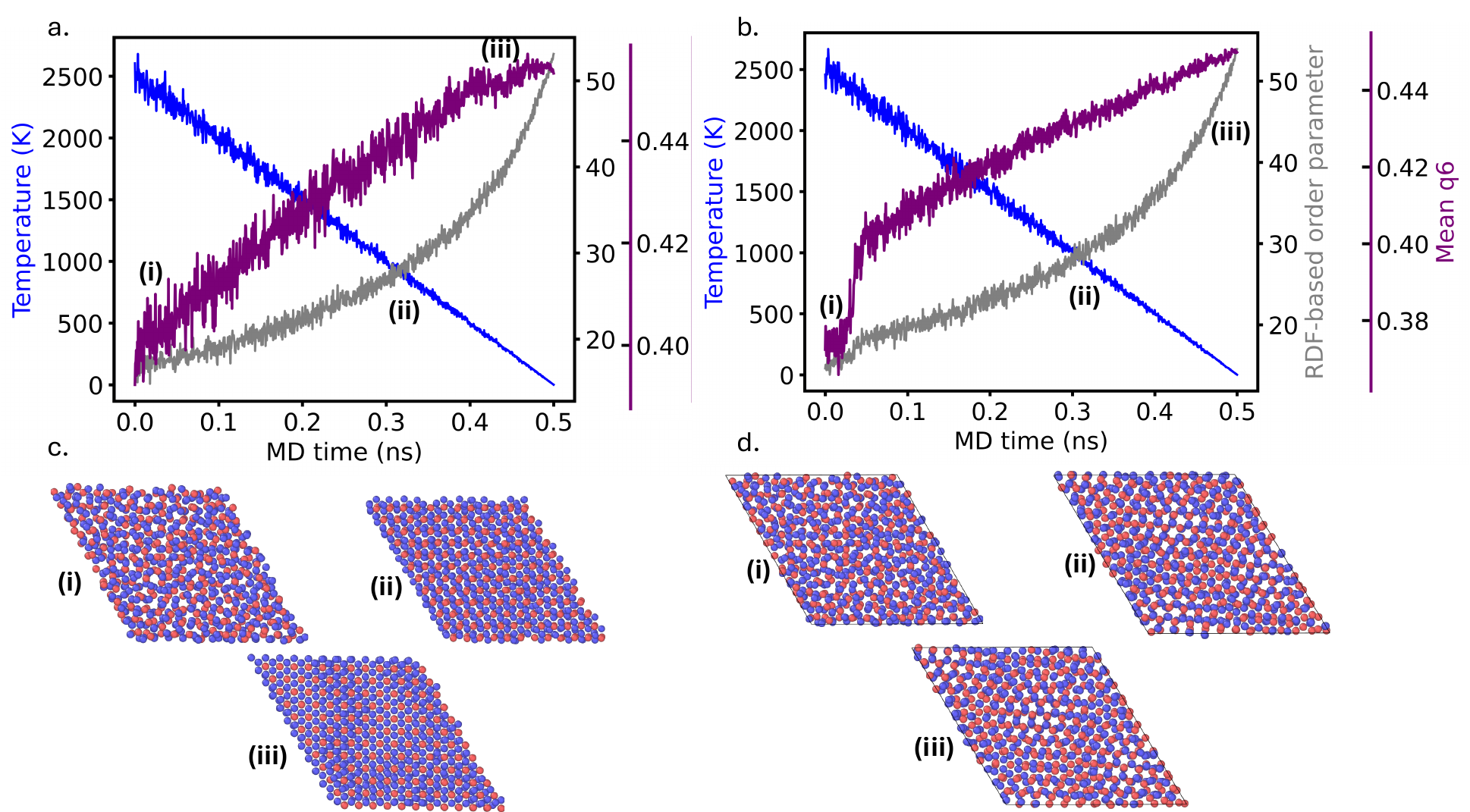}
\caption{\textbf{Recrystallization under temperature and strain effects} As temperature is quenched near 0 K \textbf{(a, b)}, overall order starts to emerge in the top layer \added{as shown by both the mean-per atom $q6$ (purple) and pair-count based order parameters (grey) in the bottom panel \textbf{(c,d)} for different snapshot in the MD trajectory}. While only temperature driven (i.e., heating at 2500 K) trajectories lead to almost no significant interlayer twist---AB stacked 2H phase (\textbf{c-iii}), applying a lattice strain ($\epsilon_{lat}$) with a supercell shear of ($\epsilon_{sh}$) 1.5 \% each, can induce a recrystallized bilayer with a twist angle ($\theta$) of around $8.7^\circ$ (\textbf{d-iii}).}
\label{fig:recryst_traj}
\end{figure*}

Our hpc resource-driven workflow to enable accelerated computational autonomous synthesis is demonstrated in Figure \ref{fig:workflow_schem}. Choosing recrystallization of an amorphous Molybdenum disulfide (MoS$_2$) layer on top of a monolayer crystalline substrate (Figure \ref{fig:workflow_schem}a) as a case study (amorph+crystal), we aim to predict the parameter space (Figure \ref{fig:workflow_schem}c) which could transform into crystalline bi-layers with target interlayer twist angles (Figure \ref{fig:workflow_schem}e). We hypothesize, that for a given stoichiometry, the parameters space for the recrystallization process involves annealing temperature ($T$), in-plane lattice mismatch strain ($\epsilon_{lat}$, assuming it to be tensile for the thin-film) and shear ($\epsilon_{sh}$), and we could identify the appropriate growth regime leading to a specific target moir\'e structure (Figure \ref{fig:recryst_traj}). Choice of these parameters are motivated from the experimentally reported investigations of a variety of transformation and reconstruction pathways involving van der Waals layered epitaxial phases \cite{bianchini2020interplay, weston2020homo_layer_TMD_reconstruction,baek2023thermally, li2021phase}. While several other kinetic and thermodynamic driving forces e.g., chemical potential, electron beam effects, effect of edge crystallography and pressure etc. will play crucial roles in the emergence of long-range order during recrystallization and as well dictating the final orientation of phases, for practicality, we restrict ourselves to implicitly incorporate such effects through annealing temperature, hydrostatic (e.g. uniform lattice mismatch strains) as well as symmetry distortions i.e. shear components in the mechanical strain tensor to demonstrate the potential of such computational investigations in the autonomous discovery campaigns. We note that by considering this parameter space, we only demonstrate a representative workflow where a diverse set of local environments resulting from temperature and strain effects are covered during the amorphous to crystalline phase transformation dynamics; further consideration of drivers like chemical potential and pressure could also be similarly used to explore the structural evolution pathways relevant for typical growth conditions within our current framework.    

In principle, this involves exploration of materials dynamics over a nontrivial landscape sensitive to small perturbations in temperature and strain. Accelerating such search will require adaptive sampling techniques. We take a multi-level iterative acceleration approach by adopting i) Bayesian optimization (BO) to predict  along with ii) an on-the-fly asynchronous high-throughput sampling orchestrated on a leadership class single batch computational allocation (c.f. \hyperref[subsec:method-ensemble-workflow]{methods}) bypassing scheduling bottlenecks. Such workflows, hence, ensure an integrated gain in terms of optimal resource usage and computational throughput though uncertainty-aware dynamic task execution based sampling strategies \cite{bagchi2025matensemble}. 

Active learning of molecular dynamics (MD) trajectories performed at a candidate parameter combination set ($p_i \equiv \{T_i, \epsilon_{lat\ i}, \epsilon_{sh\ i}\}$, where $ i \in \{1, 2, \dots, N_{\text{batch}}\}$) also requires automated parsing through the large volume simulation output as well as analytics related to the quantity of interest (q.o.i) which in this current case is the interlayer twist angle. Our solution to address this problem is to analyze simulation data at a regular time interval on-the-fly as soon as the data becomes available through an in-memory message passing and data-parallel processing approach (c.f. \hyperref[subsec:method:on-the-fly_twist]{Methods}). Hyperparamters of this adaptive workflow, e.g., batch size of candidate parameters ($n_{batch}$) to be evaluated with MD and the available allocation are correlated (for instance, with the allocation required MD evaulation) and are managed by adaptive scheduling so as to maintain a uniform number of tasks running during most part of the wallclock limit. The capability (Figure \ref{fig:workflow_schem}) to combine on-the-fly analytics of individual simulations with a concurrent asynchronous evaluation framework makes it possible to expand the global search horizon while exploiting uncertainty of the data through BO. 

\subsection{Tunable twisted MoS$_2$ bilayers through recrystallization}
We start from an initially disordered top layer (amorph-MoS$_2$) generated with melt-quench MD runs (c.f. \hyperref[subsec:MD-recryst]{Methods}) placed on top of 2H crystalline layer (Figure \ref{fig:recryst_traj}c,d).  Upon rapid heating (e.g., $T$=2500 K in Figure \ref{fig:recryst_traj})followed by a slow quench until $\sim 0$ K, recrystallized MoS$_2$ top layer forms an AB stacked 2H bi-layer phase in Figure 
\ref{fig:recryst_traj}a with gradually emerging order (c.f. \hyperref[subsec:MD-recryst]{Methods} details on order quantification using  a pair-distance count based simplistic order parameter (gray lines in Figure 2) which compares well with a conventional hexatic symmetry identifier like $Q_6$ shown in purple in Figure 2). The final ordered structure is drastically altered as 1.5\% strain in terms of both the lattice strain ($\epsilon_{lat}$) and shear ($\epsilon_{sh}$) is induced (Figure \ref{fig:recryst_traj}b). A moir\'e bilayer with a final interlayer twist angle of $8.7^\circ$ with a supercell wavelength of $20.8 \si{\angstrom}$ appears as a result of this change in the recrystallization conditions. 
This indicates that there exists a promising approach to bias recrystallization pathways towards tunable twisted bi-layers as final phases with the help of controlling temperature and strain induced parameters. Based on such observations as well as guided by past computational investigations of melt-quench crystallization analysis \cite{ponomarev2022new} of MoS$_2$ involving reactive force-fields (c.f. \ref{subsec:MD-recryst}modeling details) we pick a range of 1500K-3000K for annealing temperature and lattice strains upto 5\% for our autonomous search space.
We note that using such approaches to control solid-solid phase transitions via self-organization of disorderd 2D materials can emerge as cleaner and tunable alternates over traditional CVD driven growth of twisted-bilayer \cite{xu2024reconfiguring} as they involve complex interplay of nucleation and growth kinetics dominated by chemically sensitive driving factors.  

\begin{figure*}[hbt!]
\centering
\includegraphics[height=0.45\linewidth,width=1.\linewidth,angle=0]{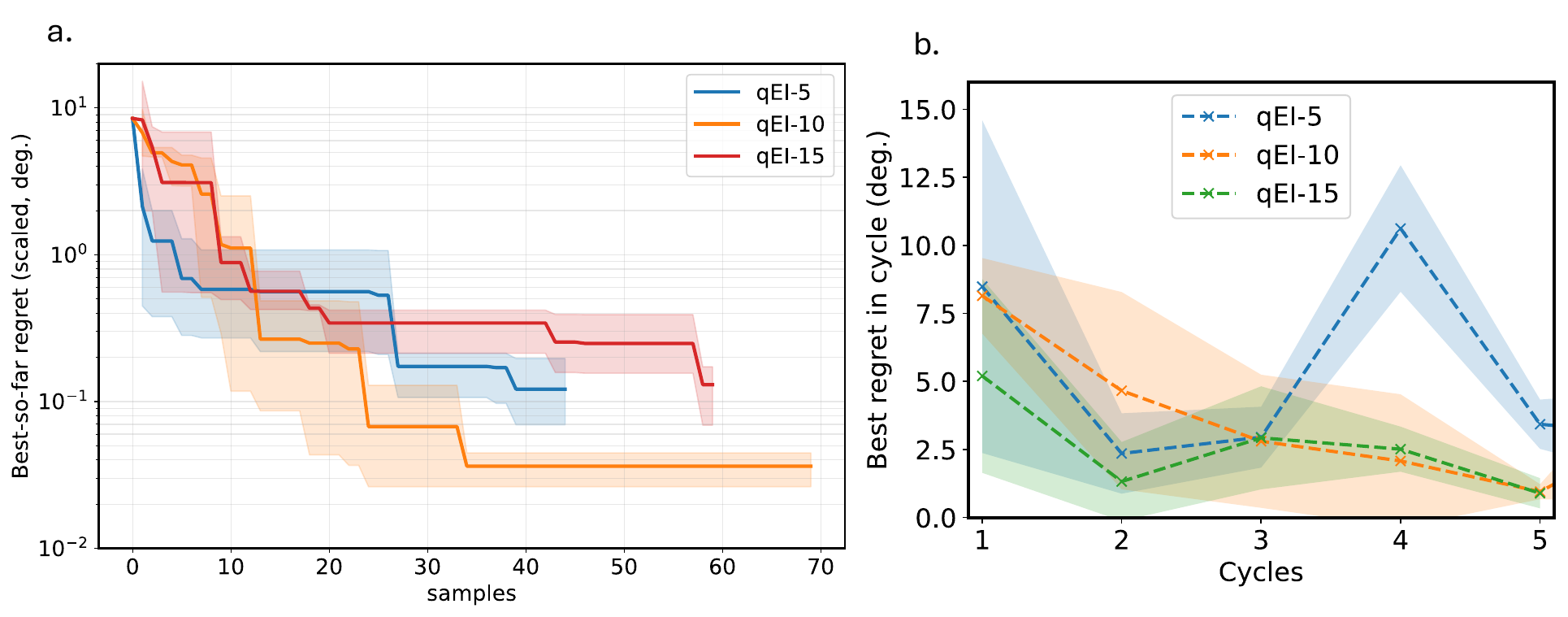}
\caption{\textbf{Sampling a target twist with batch acquisition based on q-Expected Improvement} Regret (absolute error) between sampled twist angle and the predefined target for q-EI based acquisitions with different batch sizes ($n_{batch}$) in \textbf{(a)}. \added{Solid lines correspond to the cumulative minimum of regret over the iterations. Mean regret evolution in \textbf{(b)} for the best sample in each batch over BO cycle iterations for varying batch sizes. The confidence interval shows standard deviation over 3 repetitions}} 

% }
\label{fig:qEI}
\end{figure*}

\subsection{Bayesian batch acquisitions to sample a target twist}
Given that we have a tunable epitaxial phenomenon to generate twisted homobilayer which can be captured through atomistic simulations, we now turn towards an autonomous approach to expedite the process of discovering optimal parameter combinations leading to a target twisted final state, given \textit{on demand}. While sequential acquisition based Bayesian optimization loops have been commonly deployed in self-driving laboratories \cite{ament2021autonomous,sheng2024autonomous, kusne2020fly, shields2021bayesian}, in the context of building virtual environments, throughput in terms of both sampling and decision making can be increased through extreme scale computational resources. Furthermore, many of the state-of-the art materials simulation capabilities are highly optimized into applications exploiting leadership class supercomputers \cite{alexander2020exascale}. 

Our approach therefore is to accelerate the discovery of recrystallization parameters through asynchronous iterative evaluations of a batch of candidates ($p_i,\ i=1,2 ...n_{batch}$) recommended by BO. Several algorithms have explored the idea of parallel optimization of the acquisition function with various approximation \cite{hernandez2017parallel, kandasamy2018parallelised} to account for the uncertainties of the pending evaluations. First, we start with q-expected improvement (EI) (c.f. \hyperref[subsec:method:async_BO]{Methods}) as our acquisition function to sample a target twist angle, $\theta_{\text{target}}=8.5^\circ$. In Figure \ref{fig:qEI}a, we show the variation in predictions with different batch sizes ($n_{batch}$).
The regrets from the target for each of the acquisition events are shown in Figures \ref{fig:qEI}a and \ref{fig:comparison}a. These acquisitions are driven by Bayesian uncertainty quantification metrics as each of the MD trajectory is spawned with a set of annealing temperature, lattice and shear strains under recommendation of algorithms as defined in the methods section. To better follow the progress of this regret minimization process, we added cumulative minimum lines for each case (c.f. Figures \ref{fig:qEI}a and \ref{fig:comparison}a). From this analysis we can see that the cumulative minimum gradually drops discretely as sample trajectories are evaluated. For qEI driven acquisition, with a batch size=5, the best regret ($0.1^\circ$) in Figure \ref{fig:qEI}a leads a final average recrystallized moiré twist of $8.4^\circ$, i.e. Target ($8.5^\circ$) – regret ($0.1^\circ$) as achieved on the 38$^{th}$ recommendation (5-th cycle )at a temperature of 1386 K, 1.2\%  lattice strain and 5.53\% shear (Supplementary Figure \ref{fig:SI-sampling_evol}).

Whereas, starting with an identical set of initial observation data, with a larger batch size ($n_{batch}$=15), the workflow predicts a twisted phase closer to the target, $\theta=8.46^\circ$ (Figure \ref{fig:qEI}a and Supplementary Figure \ref{fig:SI-sampling_evol}c,d) for temperature=2077 K, 2.3\% lattice strain and 5.76\% shear at 4-th cycle (58-th sample). 

\begin{figure*}[hbt!]
\centering
\includegraphics[height=0.4\linewidth,width=0.95\linewidth,angle=0]{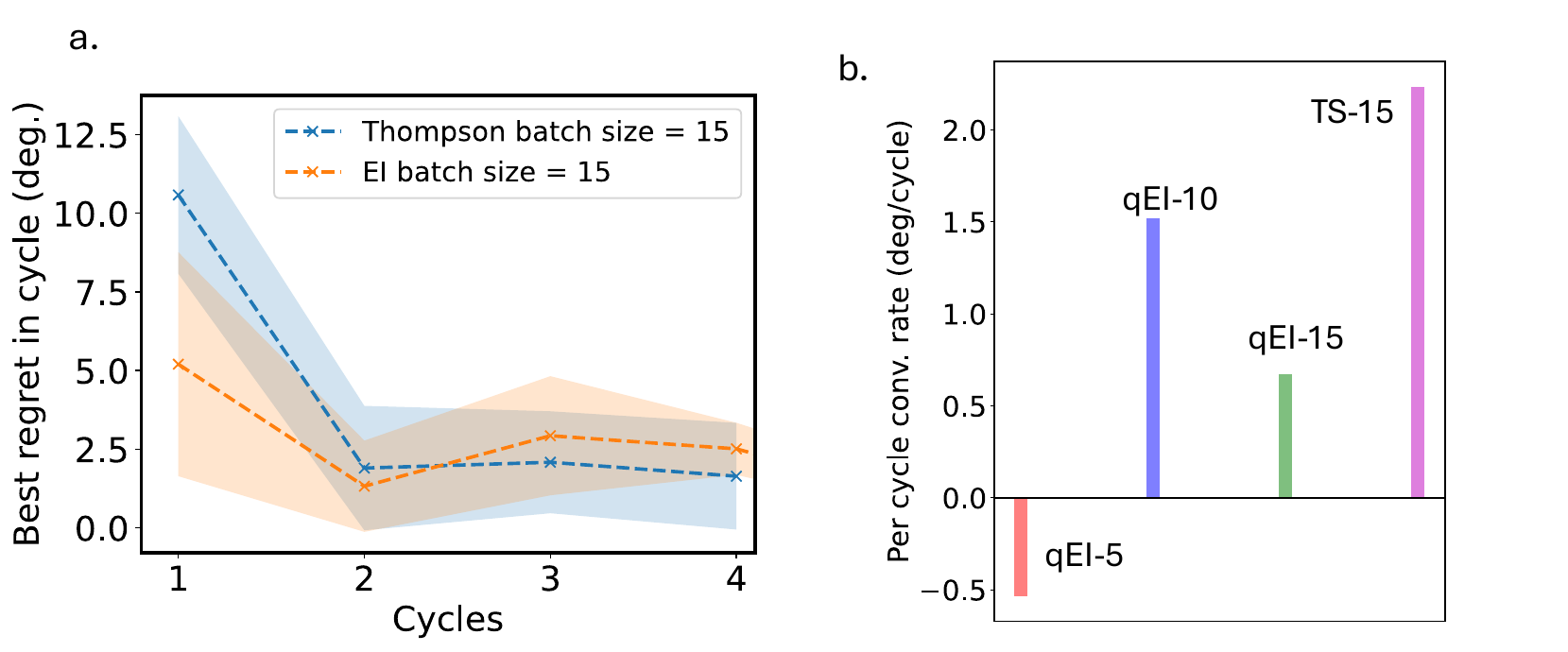}
\caption{\textbf{Effect of batch size and sampling strategies} Initialized with identical set of observations, performance of two different acquisition strategies with batch evaluation--Thompson sampling and Expected Improvement in \textbf{(a)}. Overall efficiency of various batch sampling strategies with respect to size and acquisitions in \textbf{(b)}. Inherent stochastic nature of Thompson sampling outperforms EI based acquisition which can become explorative beyond an optimal batch size}
\label{fig:TS_async_eff}
\end{figure*}

While it is obvious that larger batches can be harnessed to cover a larger search space  (Supplementary Figures \ref{fig:SI-sampling_evol}b,d) in faster wallclock time with scalable resources, the indirect exploitation of the posterior, as it is the case for EI as an acquisition function (\cite{gonzalez2016batch}), might not always lead to efficient searches (Figure \ref{fig:TS_async_eff}a) with a larger batch size. We could define a \added{per-cycle} sampling efficiency metric as an average rate \added{(computed normalizing the change of mean best regret per cycle over 4 cycles)} describing how fast BO iteration cycles generate an MD trajectory with a target twisted phase. As seen in Figure \ref{fig:TS_async_eff}(b),  this is driven by the fact that beyond a point larger batches tend to make the search more explorative rather than efficiently exploiting model uncertainties accessed through a implicit acquisition criteria like EI \cite{de2019sampling}.

\begin{figure*}[hbt!]
\centering
\includegraphics[height=0.4\linewidth,width=1\linewidth,angle=0]{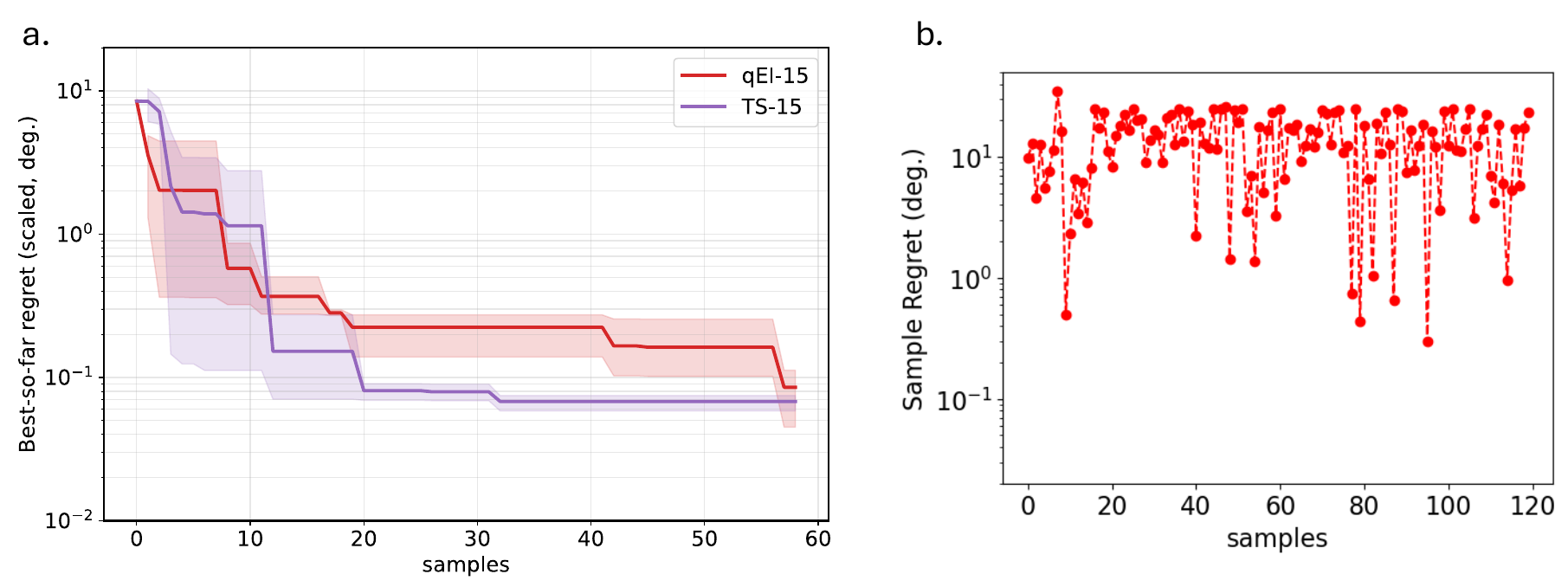}
\caption{\textbf{Comparison of adaptive batch sampling via BO with random sampling} Performance of two different acquisition strategies with batch evaluation--Thompson sampling and Expected Improvement \textbf{(a)}.  Solid lines correspond to the cumulative minimum of regret over the iterations. Initialized with identical set of observations, random sampling \textbf{(b)} underperforms within a comparable sampling duration.}
\label{fig:comparison}
\end{figure*}
To probe the effect of acquisition strategies, we adopt batch Thompson sampling (c.f. \hyperref[subsec:method:async_BO]{methods}), shown in Figure \ref{fig:comparison}a, for the same target angle ($\theta=8.5^\circ$) and identical prior observation set. It is interesting to note that Thompson sampling with a comparable batch size (i.e., $n_{batch}=15$) leads to a more consistent sampling near the target yielding three twist angles (sample index) e.g., ${8.39^\circ (18),  8.57^\circ (47)}$ at different parameter combinations ($T_{18}=1362 \text{K}, \epsilon_{lat \ 18}=0.4\%, \epsilon_{sh \ 18}=5.69\%$), ($T_{47}=3195 \text{K}, \epsilon_{lat \ 47}=0.2\%, \epsilon_{sh \ 47}=5.5\%$) than EI. This ability to scan through a diverse range of recrystallization parameters could be due to the probabilistic nature of Thompson sampling from the posterior leads to a better exploration-exploitation trade-off in batch mode \cite{gonzalez2016batch} than primarily indirect exploitation (of posterior properties) driven acquisitions through EI. Furthermore, we assess the advantage of our active learning-driven sampling approach over vanilla random sampling (Figure \ref{fig:comparison}b). Starting with the identical initial observation set like all the previous sampling demonstrations, random sampling with $n_{batch}=15$ is unable to sample twist angles close to the target ($8.5^\circ$) even upto 120 samples, with the best value being $8.2^\circ$. 
% Although in this study there is surrogate model mismatch with priors, AL has proven better performing overall than random sampling as has been shown in previous studies \cite{Lookman2019}. In this framework we have defined the \textit{stopping criterion} for the AL based on the ability of the Bayesian optimization (BO) loop to consistently sample interlayer twist angles within a tolerance of $\sim$ $1 \times 10^{-1}$ degrees) [from predefined twist angle- value??] This tolerance is deemed appropriate based on previous studies... \cite{}.

\begin{figure*}[hbt!]
\centering
\includegraphics[height=0.55\linewidth,width=1.\linewidth,angle=0]{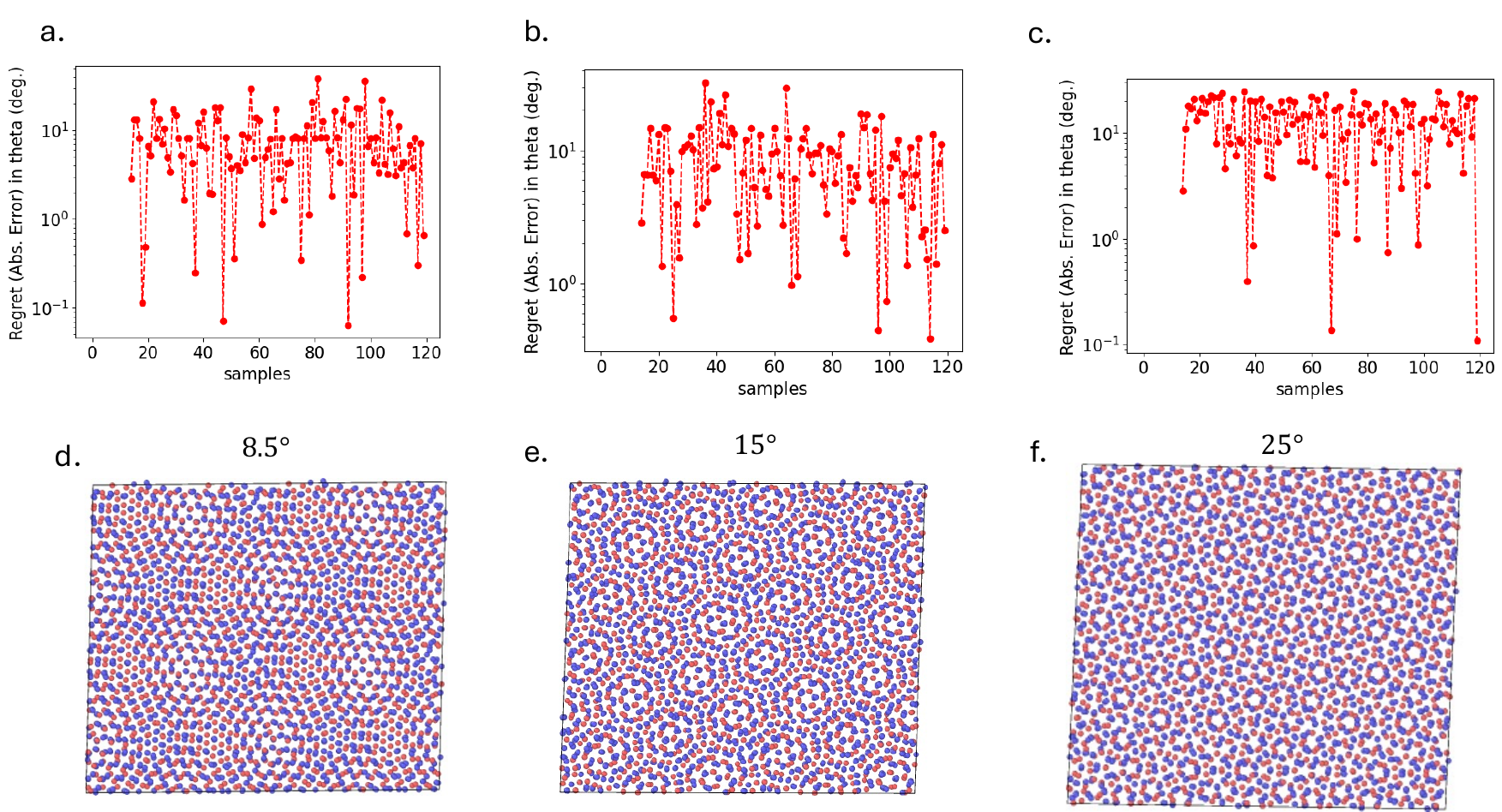}
\caption{\textbf{Achieving recrystallization with \textit{on demand} final twist angles} General flexibility of the workflow is shown using batch Thompson sampling. Starting from the same set of initial samples, the asynchronous Thompson sampling could search three diverse set of moir\'e sublattices with target interlayer twists--$8.5^\circ$ \textbf{(a,d)}, $15^\circ$ \textbf{(b,e)} and $25^\circ$ \textbf{(c,f)}.}
\label{fig:on-demand}
\end{figure*}

\section{Discussions}
\subsection{Designing twisted bilayer interfaces \textit{on-demand}}
We thus far have demonstrated a capability to identify target recrystallization conditions through our Bayesian optimization workflow coupled with asynchronous ensemble evaluation strategies driven by scalable computational resources. While constrained in the scope of parameter space (e.g., more realistic conditions might involve more drivers like pressure, chemical potential, rate of heating etc.), these results show the potential of accelerating targeted solid state phase transitions in vdW 2D materials interfaces. To further demonstrate the efficacy of our workflow we provide quantitative evidence towards enabling \textit{on-demand} design of twisted moir\'e interfaces of MoS$_{2}$ bilayers. Using batch Thompson sampling as a mode to evaluate the parameter space comprising temperature, lattice strain and shear, we focus  our workflow towards sampling twist angles $15^\circ$ and $25^\circ$ in addition to $8.5^\circ$--thereby covering a wide range of moir\'e wavelengths and superlattices. As shown in Figure \ref{fig:on-demand}, the current approach is able to achieve recrystallization into moir\'e lattice with predefined target twist angles, especially for structures with nontrivial target twist angles (e.g. $25^\circ$) (c.f. Supplementary Figure \ref{fig:SI-hist} for more details). 

\added{It will be further intriguing to probe the effects of edge reactions, stacking orders, perturbed local
environments and defect-pinning/doping that results from atomic rearrangement during the melt-
quench recrystallization as well as precursor-driven growth \cite{filho2024density}, on their electronic structure via first principles. The advantage of proposed online
workflow orchestration strategy is that such effects can now be more readily investigated leveraging
HPC platforms.}

Moreover, we successfully identify multiple combinations of parameter corresponding to for each of these diversely specified target twists (Figure \ref{fig:on-demand}). For instance, interlayer twists close to $15^\circ$ can be recrystallized at i)2725 K, 2.7\% lattice strain and 5.5\% shear as well as at ii) 2107 K, 1.3\% lattice strain with 5.69\% shear. Likewise, we identify more than one set of optimal parameters for $\theta\sim 25^\circ$---i) \{2797 K, 0.9\% lattice strain and 5.57\% shear\} and ii) \{1379 K, 3.2\% lattice strain and 5.51\% shear\}. We note that very small changes to the recrystallization parameters, especially shearing the supercell leads to drastically different final moir\'e superlatices. For this reason, we keep the search window for $\epsilon_{sh}$ to vary from 5.5\% to 5.75\% strain levels which corresponds to absolute distances of $1 \si{\angstrom}$ to $1.5 \si{\angstrom}$ tilts of in-plane supercell dimensions. Furthermore, the wide variations (Supplementary Figure \ref{fig:SI-param_diversity}) in optimal temperature and lattice strain conditions associated with a specific final twisted bi-layer phase, leads us to speculate that the recrystallization landscape could be rough having multiple locally varying optimal cases. Traditional grid search or sequential active learning protocols hence might not be adequate to exhaustively identify favorable recrystallization conditions for a given target twist (Supplementary Figure \ref{fig:SI-twist_landscape}). 
Furthermore, current approach of high-throughput classical force-field driven simulations being linear scaling O(N) in comparison to cubic scaling ab initio/DFT simulations O(N$^3$), making them much more competitive towards accessing sufficient timespan for the \textit{in-silico} recrystallization simulation recipe describe here, within a matter of few hours of wallclock time with hpc-scalable batch parallel BO as long as enough compute is available. 
However, we note that bridging the timescales of state of the art atomistic simulations with what is relevant for experiments is a long standing multiscale challenge, for which we remark that our end-to-end developments of asynchronous on the fly active learning atomistics are ideally suited to upscale the dynamics with coarse grained models e.g., kinetic monte carlo, phase field simulations etc. \added{The presented online ensemble sampling workflow for identifying recrystallization regimes of MoS2 bi-layers could be seamlessly extended towards various other TMDC systems (.e.g, W/Mo-S/Se/Te etc.), multilayers or even heterostructures are of increasing interest due to their highly tunable quantum properties. In practice, such adoptions are not limited by any of the workflows features which are rather general and mostly take atomistic systems and force-field as inputs. If appropriate force-field descriptions are available for a TMDC system of choice, the current framework should be fully transferable. With the rise of highly accurate foundation MLIPs, the bottlenecks of force descriptions are rapidly disappearing.  Adapting these methods to more stochastic or continuum simulation approaches is also quite straightforward.}

\section{Conclusion}

We have demonstrated that combining extreme-scale computations, Bayesian optimisation and asynchronous automated computational workflows, a powerful building block 
 to enable \textit{on-demand} synthesis can be achieved, and that it allows time and resource efficient discovery of synthesis pathways for complex materials.  The targeted problem of achieving `on-demand' van der Waals epitaxy is a major technological challenge, specifically when controlling both the phase and the interlayer stacking orientations.  We demonstrate that within ~20 iterations the autonomous computational synthesis platform achieves the targeted moiré quantum heterostructure. The global sampling opportunities boosted by extreme-scale resource-driven batch acquisition workflows and on-the-fly analysis and UQ capabilities are ideally suited to pin down more sophisticated interplay of thermodynamic and kinetic factor driving self-organized and strain-engineered interfacial evolution during vdW epitaxy.  Furthermore, we note that integrating the autonomous computational synthesis workflows with real experimental synthesis efforts can quickly narrow down the number of experimental attempts to grow a targeted quantum heterostructure.  The current ensemble simulation framework can also be used to generate digital-twins on experimental measurements, such as X-ray diffraction (XRD) and reflective high-energy electron diffraction (RHEED). As such, integration of our framework with experimental molecular beam-epitaxy (MBE) or pulsed-laser-deposition (PLD) systems will allow us to perform on-the-fly comparison with experimental XRD/RHEED measurements, so as to guide experiments in real-time, allowing a real autonomous thin-film synthesis platform with theory-in-the-loop.

\section{Methods}
\subsection{\texttt{MatEnsemble}: Adaptive real-time ensemble task management environment for extreme scale computing}\label{subsec:method-ensemble-workflow}

Within a single HPC batch allocation, ensemble evaluations/jobs which require MPI parallel resources at the individual level, often suffer from scheduling challenges with typical SLURM-based workload managers \cite{yoo2003slurm}. This is primarily due to the fact that a hard ceiling (e.g., 100 for Frontier, Perlmutter/NERSC) are often imposed toward chaining together multiple job submissions in batch mode (e.f. \texttt{srun} for SLURM). Such setups are highly unsuitable when \textbf{(i)} size of the ensemble exceeds the ceiling of a scheduler and  \textbf{(ii)} dynamic heterogeneous tasks are at the core of the workflow e.g., based on outcomes of the ensemble of MD simulations active learning or UQ algorithms (e.g., BO) has to operate and further spawn next generation of ensembles. 

To mitigate the above challenges, we implement all our active learning MD simulations within a recently developed adaptive and dynamic orchestrator framework \texttt{matEnsemble} \cite{bagchi2025matensemble} which has an executor back-end of \texttt{Flux}, an Exascale friendly HPC resource manager with hierarchical graph-based scheduler allowing for a generalized flexible \textit{custom} operations within a single batch allocation.
\texttt{MatEnsemble} \cite{bagchi2025matensemble} benefits from the native python executor-interface of \texttt{Flux} \cite{ahn2020flux}, and the concurrent asynchronous programming model of core python through \texttt{Future} objects \cite{quinlan2009futures}. A continuous throughput is maintained via dynamically spawning and monitoring task. Furthermore, to enable real-time streaming of post-processed data (e.g., interlayer twist angle evolution) from large-scale atomistic trajectories an \textit{in-memory} data analysis protocol is used by exploiting the heterogeneous (GPU+CPU) architecture of Exascale systems (e.g., Frontier) via a round-robin MPI-communicator splitting approach (c.f. \cite{bagchi2025matensemble}). As explained in the following sections, such an online adaptive framework enables efficiently coupling between available computing resource chunks for ensemble evaluations guided by adaptive sampling methods.

\subsection{Parallel batch sampling with Bayesian optimization}
\label{subsec:method:async_BO}
Traditionally optimal design of experiments via active learning based on Bayesian optimization \cite{movckus1975Bayesian} proceeds through sequential acquisitions. With the adaptive asynchronous capability to produce scalable ensemble evaluations in our workflow infrastructure, batch mode acquisitions algorithms are natural choices to effectively utilize available hpc resources and accelerate the search towards target twist angles in the spirit of efficient global optimization \cite{jones1998efficient} with the following acquisition functions.  

\noindent \textbf{q-Expected Improvement:}
With a goal to accelerate the solution of the minimization problem for an expensive black-box function $f$, we have used parallel/batch expected improvement\cite{Daulton23, ginsbourger2010kriging, Kushner1964ANM, Ginsbourger2008AMC} in this study. It is computed by Monte Carlo sampling and is given by:
\begin{align}
\mathrm{qEI}_{y^{*}}(\mathbf{X}) &= \mathbb{E}_{f(\mathbf{X})}\left[\max _{j=1, \ldots, q}\left\{\left[f\left(\mathbf{x}_{j}\right)-y^{*}\right]_{+}\right\}\right] \\
                &\approx \sum_{i=1}^{N} \max _{j=1, \ldots, q}\left\{\left[\xi^{i}\left(\mathbf{x}_{j}\right)-y^{*}\right]_{+}\right\}
\end{align}
\noindent where $\xi^{i}(\mathbf{x}) \sim f(\mathbf{x})$ are samples drawn randomly from the joint posterior distribution of the model evaluated at batch points $\mathbf{x}$, before optimizing the expectation which dictates the acquisition strategy rather than inferences which are directly based on the posterior, and top performing $n_\text{batch}$ number of candidates from the optimization are finally obtained. Relevant to the last point, we also explored more direct posterior sampling strategies e.g. Thompson sampling as described in the following.

% which is the absolute distance between sampled twist angle ($\theta$) and target twist $\theta_{\text{target}}$, such that
% \begin{equation}
%     x^{*} = \arg \min_{\mathbf{x} \in \mathcal{X}} f(\mathbf{x})
% \end{equation}

% where $\mathcal{X}$ is input parameter space with $N$ candidates $\{x_i\}_{i=1...N}$ with $x_i \equiv \{T_i, \epsilon_{lat}, \epsilon_{sh}\}$.  Assuming a $\mathcal{GP}(x)$ surrogate approximates $f$, standard (sequential) Expected Improvement acquisition is defined as 
% \begin{equation}
%     a_{\text{EI}} (x | \mathcal{D}_t) = \mathbb{E}[\max\left(f^*(x) - f(x), 0\right)]
% \end{equation}

% for until $t$-cycles of observations ($\mathcal{D}_t=\{x_i, f(x_i) \ | i<t\}$) where $f^*(x)$ is smallest objective value in the current set. The next point $x_{t+1}$ is chosen such that 
% \begin{equation}
%     x_{t+1} = \arg \max_{\mathbf{x} \in \mathcal{X}} a_{\text{EI}}
% \end{equation}

% For an analytical closed form expression of the above, c.f. \cite{Kushner1964ANM, Ginsbourger2008AMC, ginsbourger2010kriging}. 

% A batch parallel/multipoint version of the standard EI acquisition, known as \textit{q-EI} leads to a modified improvement acquisition to be maximized to obtain $x_{t+1}$, 
% \begin{equation}
%     a_{\text{q-EI}}=\mathbb{E}\left[ \sum_{i=1}^{n_{\text{batch}}} \max(f^* - f(x_i), 0) \right] \approx \sum_{i=1}^{n_{\text{batch}}} \mathbb{E}\left[ \max(f^* - f(x_i), 0) \right]
% \end{equation}

\noindent \textbf{Batch Thompson sampling:}
Unlike several acquisition functions which are mostly driven by exploitative strategies, Thompson sampling, due to its stochastic nature, is able to handle the exploration-exploitation trade-off in a better fashion. In our simulation we adopt the batch parallel version of Thompson sampling first proposed by Kandasamy et al. \cite{kandasamy2018parallelised}. We choose a quasi-Monte Carlo sampler which uses Sobol sequences \cite{snoek2012practical} to sample from the joint-posterior of over q ($n_{\text{batch}}$) batch and optimizes over simple regret \cite{gonzalez2016batch}, 
\begin{equation}
  \text{Simple Regret}(\mathbf{X}) = f^* - \min_{i = 1, \dots, n_{\text{batch}}} f(\mathbf{x}_i)  
\end{equation}
The optimization leads to $n_{\text{batch}}$ number of top candidates to be evaluated in the following cycle. Upon finishing the evaluation of all candidates in the batch the next cycle of the BO is resumed.

\noindent We implement all our BO algorithms using the \texttt{BOTorch} library \cite{balandat2020botorch} for single task Gaussian process approximated with Mat\'ern kernel \cite{williams2006gaussian}.

\subsection{Molecular dynamics simulations of recrystallization in MoS$_{2}$ bilayers}\label{subsec:MD-recryst}
We perform all our MD simulations by wrapping the shared library utilities of the open source package \texttt{LAMMPS}\cite{thompson2022lammps} through \texttt{matEnsemble} drivers. Starting with a 2H bilayer MoS$_2$ phase (rhombohedral supercell with 864 atoms), we first induce disorder in the top crystalline layer by selectively heating up the layer at a high temperature of 5000K for 50ps with a timestep of 0.5fs. We use a classical reactive force field ReaxFF, specifically re-parametrized to accurately capture order-disorder transitions in MoS$_2$ during multiple melt-quench cycles \cite{ponomarev2022new}. After the top layer the undergoes melting, the whole bilayer system is equilibrated at 300K in atomospheric pressure using a Berendsen NPT thermostat for 50ps to achieve the disordered top layer placed on top of a crystalline bottom layer---this served as the initial sample (amorph+Xtal) for the recrystallization simulations. 

% and orthogonal box with ~20,660 and 0.3 million atoms

Finally to simulate the recrystallization process under various candidate input parameter combination of $p_{i} \equiv \{T_i, \epsilon_{lat}, \epsilon_{sh}\}$, the system is rapidly heated for 50ps at a temperature $T_{i}$ along with in-plane lattice rescaling ($x$, $y$) of $(1+\epsilon_{lat})$ (to model epitaxial strain) and an in-plane shear strain ($xy$) of $\epsilon_{sh}$ measured on the substrate crystal layer (c.f. Figure \ref{fig:workflow_schem}a). This is then followed by a slow quench of the bilayer system to reduce the temperature level from $T_{i}$ to 0 K over 0.5 ns i.e. $1\times10^6$ MD steps. The final step of rapid melt-slow quench simulations are performed using Langevin thermostat (NVT). Periodic boundaries are maintained along in-plane (i.e. $x$ and $y$) directions. Lennard-Jones reflective walls are implemented near the bottom ($0 \si{\angstrom}$) and top ($12 \si{\angstrom}$) edges. We use Ovito for all the visualizations and analysis of MD trajectories.

\added{Important to note that the annealing approach does not guarantee lowest energy configurations which is intentional
for the present study to let the workflow explore high-energy metastable basins and discover
a diverse range of on-demand Moire motifs which might not necessarily be equilibrium states. In the context of synthesis and materials discovery, where tracking down non-equilibrium
phases are of particular interest, such melt-quench dynamics approach could specifically be useful. While it is possible for atoms to undergo rare diffusion event which are path
dependent mostly in the long-time slow quench limit, our 0.5ns quench protocol could still be relatively fast to favor such path dependent quenching behavior in the first order. This could
be explored in more detail by studying long-time dynamics limits which is beyond the current
scope.}

% \subsubsection{Order-disorder analysis:} 
\textbf{Constructing an order parameter}--To quantify the emerging order in the MoS$_{2}$ bilayers (Figure \ref{fig:recryst_traj}), we define a simple pair correlation based custom order parameter which estimates sharpness of the first nearest neighbor peak distribution. First, the pair correlation function is computed with cutoff of $3 \si{\angstrom}$ based on first nearest neighbors. Then we count the number of pairs within a pair distance window ($[2.4 \si{\angstrom}, 2.6 \si{\angstrom}]$), where first nearest neighbor peak/mode of the distribution for pristine MoS$_2$ is around $2.47 \si{\angstrom}$.

\subsection{Computing interlayer twist-angles on-the-fly}
\label{subsec:method:on-the-fly_twist}
Adopting a streaming approach to analyze MD trajectories as they become available through a distributed in-memory message passing framework, we accelerate and perform on-the-fly data reduction. As described in more detail in \cite{bagchi2025matensemble}, apart from being scalable and ideally suited for integrated active learning workflows, such frameworks serve as general testbeds to a variety of atomistic analysis and visualization algorithms. For this study, we wrap through a simple and computationally light-weight iterative algorithm to compute the interlayer twists for atomistic snapshots along the recrystallization trajectory.  
\begin{algorithm*}
\caption{Iterative estimation of interlayer twist angles}
\textbf{Input:} atom positions $x_{i}^{j}, i \in \{1, 2, 3\} \text{ and } j \in \{1, 2, \dots, N_{\text{atoms}}\}$ and atom types $t_{j}=$ either "Mo" or "Se"\\
\textbf{Output:} Twist ($\theta_{12}$) between upper and lower layers

\begin{algorithmic}[1]
\State For each layer $l$ randomly select an atom ($j$) of a particular type (say, "Mo" )
\State Find its second nearest neighbors $\{x^k\}$ (e.g., for crystalline layers $size(\{x^k\}) \sim 6$ and $type(\{x^k\}) 
\equiv$ "Mo") 
\State Sort $\{x^k\}$ based on their in-plane polar coordinates $\{r, \theta\}^k$ with central atom $j$ position ($x^j_i$) as the origin. 
% \State Store the pair \( (\text{atom\_1}, \theta_1) \)
\State $ic \gets 0$, $\theta_{ic} \gets \min\{\theta\}^k$, $atom_{ic} \gets atom(\text{argmin}\{\theta\}^k)$ 
\While{$ic\leq \text{max iterations}$}
    \State shift central atom at $atom_{ic}$ repeat steps 2 to 3   
    \If{the neighbor selection fails due to a vacancy or other variations}
        \State Add a ghost atom at the guess position by replicating and shifting the central atom along \(r_{ic} \) in \( \theta_{ic} \)
    \EndIf
    \State $\theta_{ic} \gets \min\{\theta\}^{ic}$, $atom_{ic} \gets atom(\text{argmin}\{\theta\}^{ic})$ 
    \State $ic \gets (ic+1)$
\EndWhile

\State Fit a straight line ($y=m^lx+c^l$) over the atom positions stored in $\{atoms\}_{ic}$, $ ic \in \{1,2, \dots, \text{max iterations}\}$, where $l$ is layer number $\in \{1,2\}$

\State Finally calculate the twist $\theta_{12} \gets \arctan(\frac{|m^1-m^2|}{|1+m^1m^2|})$)

\end{algorithmic}
\end{algorithm*}
\clearpage

\added{\subsection{Analysis of polycrystallinity and validity of the twist computation algorithm}
To test the validity of our approach, we performed a more rigorous per-atom bond-orientation based interlayer twist computation approach. As can be seen (Supplementary Figure \ref{fig:single-crystalline}) from the randomly considered final quenched phases for a series of recrystallization simulations under various temperature, lattice and shear strain conditions, the twist distributions mostly tend to be \textbf{unimodal} leading to no speculation of multiple grains/polycrystalline nature. Barely some small clusters of other angles are observed which are likely due to disorder and vacancy debris after the quench. Furthermore, the \textbf{mode} of per-atom twist-histograms are with excellent match with our much simpler single twist results for most of the cases. Note that due to hexagonal symmetry, angles that are numerically close to 0 deg. and 60 deg. are physically identical--hence the bimodal distributions are still single crystal phases.}

\section{Acknowledgement}
This research is sponsored by the INTERSECT
Initiative as part of the Laboratory Directed Research and
Development Program of Oak Ridge National Laboratory,
managed by UT-Battelle, LLC, for the U.S. Department of
Energy under contract DE-AC05-00OR22725 via the QCAD project. This research used resources of the Oak Ridge Leadership Computing Facility at the Oak Ridge National Laboratory, which is supported by the Office of Science of the U.S. Department of Energy under Contract No. DE-AC05-00OR22725, via the Innovative and Novel Computational Impact on Theory and Experiment (INCITE) program.  Resources of the Compute and Data Environment for Science (CADES) was used via a user project at the Center for Nanophase Materials Sciences (CNMS), which is a US Department of Energy, Office of Science User Facility at Oak Ridge National Laboratory
% \section{Data Availability}
% Raw data generated in this work will be available from the lead corresponding author (bagchis@ornl.gov) upon reasonable request.
% \section{Code Availability}
% All codes associated with this work will be released open-source during the publication of the paper.
% % \href{https://github.com/Q-CAD/MatEnsemble/}{https://github.com/Q-CAD/MatEnsemble/}
\section{Author Contributions}
SB designed the core real-time adaptive learning workflow and on-the-fly MD simulations of recrystallization, implemented batch TS sampling and wrote the manuscript draft. AG co-wrote and designed the BO approach. AB co-wrote, implemented BO and analyzed results. PVB discussed overall results and co-wrote. PG wrote, supervised and conceived the project. 
% \clearpage
\bibliography{references}

\clearpage
\appendix
\section{\centering Supplementary materials}
\renewcommand{\thefigure}{\arabic{figure}}
\captionsetup[figure]{labelformat=simple, labelsep=colon, name={Supplementary Figure}}
\setcounter{figure}{0} 

\begin{figure}[hbt!]
\centering
\includegraphics[height=1.0\linewidth,width=0.8\linewidth,angle=0]{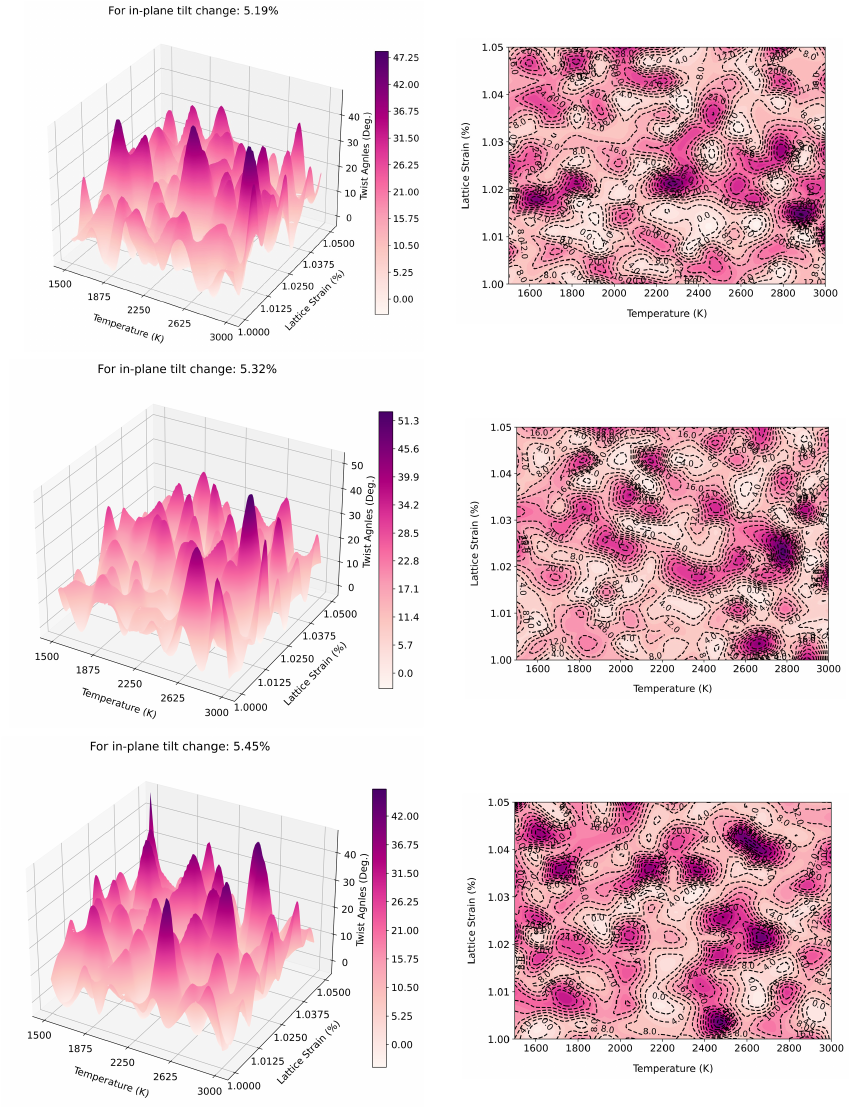}
\caption{\textbf{Exploratory landscape of twist angles observed as a function of temperature and lattice strain} extracted for various shear (in-plane tilt ratios) as depicted in the figures. The roughness of the landscape is indicative of the fact that traditional local as well as sequential Bayesian sampling approaches might perform poorly, showcasing the need for more global and stochastic asynchronous sampling strategies as explored in this work.}
\label{fig:SI-twist_landscape}
\end{figure}

\begin{figure}[hbt!]
\centering
\includegraphics[height=.4\linewidth,width=1.0\linewidth,angle=0]{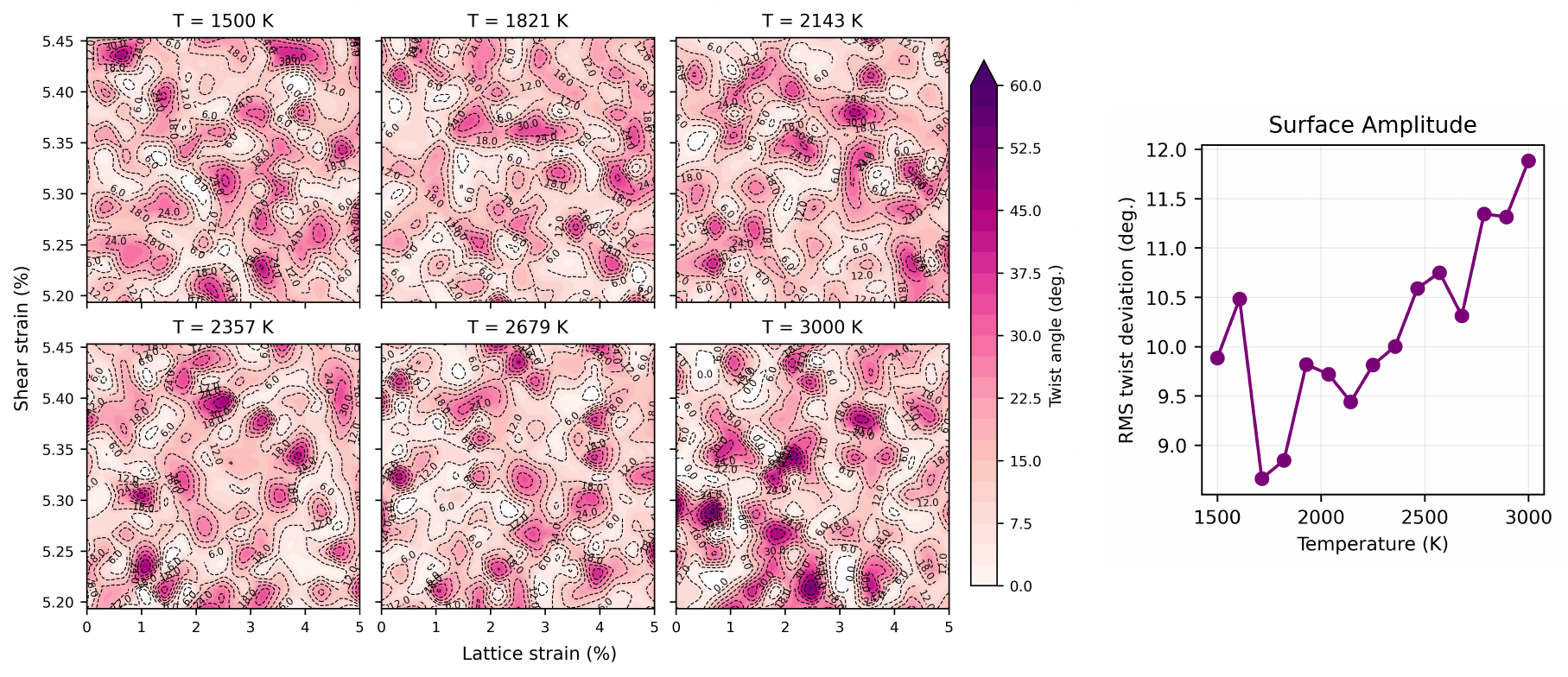}
\caption{\added{\textbf{Temperature effect on the roughness of twist-angle response landscape as functions of strain} from dataset as depicted in the previous figure. RMS of twist deviation quantified as roughness of the landscape is seen to increase with higher temperature \textbf {(b)}.}}
\label{fig:SI-temp_twist_landscape}
\end{figure}

\begin{figure}[hbt!]
\centering
\includegraphics[height=.3\linewidth,width=1.0\linewidth,angle=0]{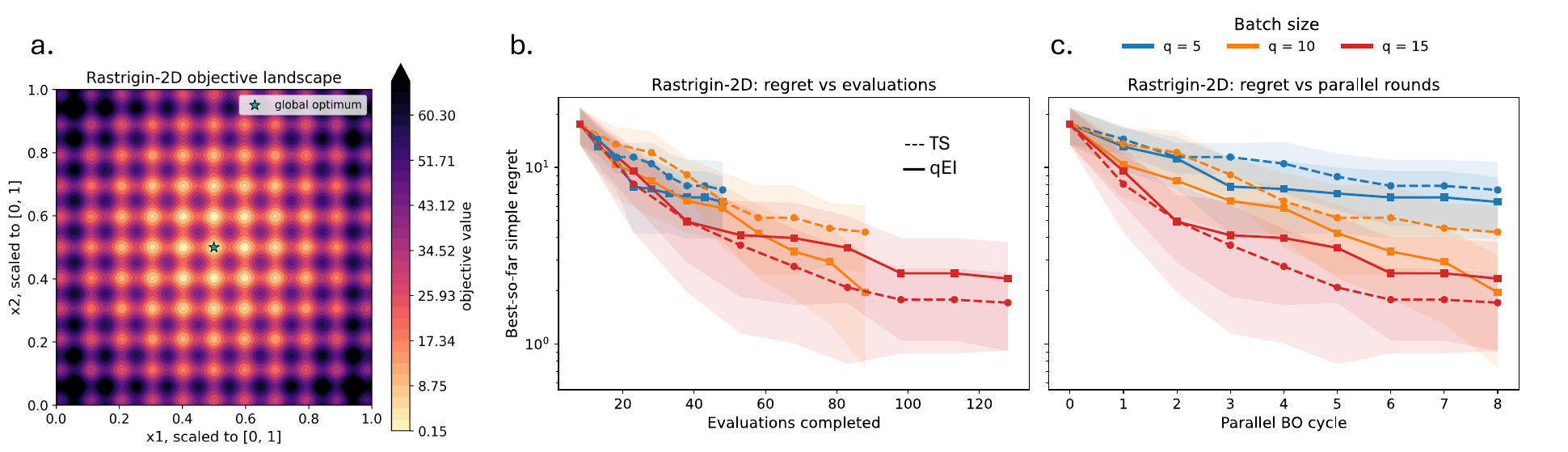}
\caption{\added{\textbf{Benchmark multimodal objective and Performance comparison of sampling strategies for a synthetic Rastrigin objective}. To test our argument on the efficiency of batch parallel EI and thompson sampling further, we investigate their performance on a synthetic benchmark multimodal objective surface described by a 2D Rastrigin function ($f(z) = 20 + \sum_{j=1}^{2}\left[z_j^2 - 10\cos(2\pi z_j)\right]$).  In the large batch limit (q=15), Thompson samples leads to better optimization in comparison to qEI with similar batch size (i.e. 15) as shown by the cumulative minimum regret vs samples (a) and vs cycles (b). Furthermore towards this end of 8 cycles, EI with a batch size of 10 is more performant than q=15. Both of these arguments echoes with our results and discussions presented in the main article.}}
\label{fig:SI-synthetic_comp}
\end{figure}

\begin{figure}[hbt!]
\centering
\includegraphics[height=.75\linewidth,width=1.\linewidth,angle=0]{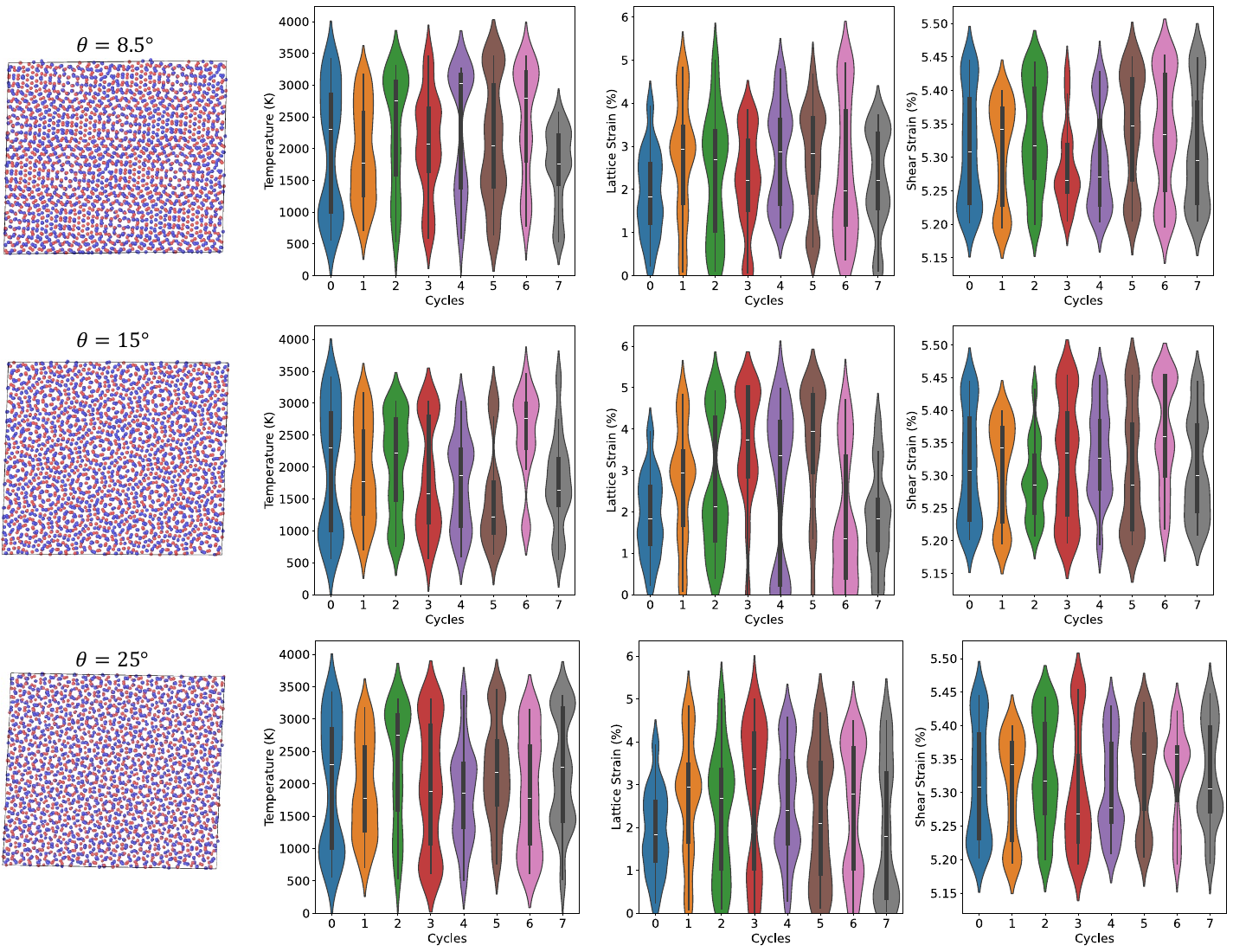}
\caption{\textbf{Diversity of growth parameters in each batch over the BO iterations} The violin plots show the ability to navigate through a diverse set of growth paramters across individual batches through Thompson Sampling with "on demand" target twists}
\label{fig:SI-param_diversity}
\end{figure}

\begin{figure}[hbt!]
\centering
\includegraphics[height=.8\linewidth,width=1.\linewidth,angle=0]{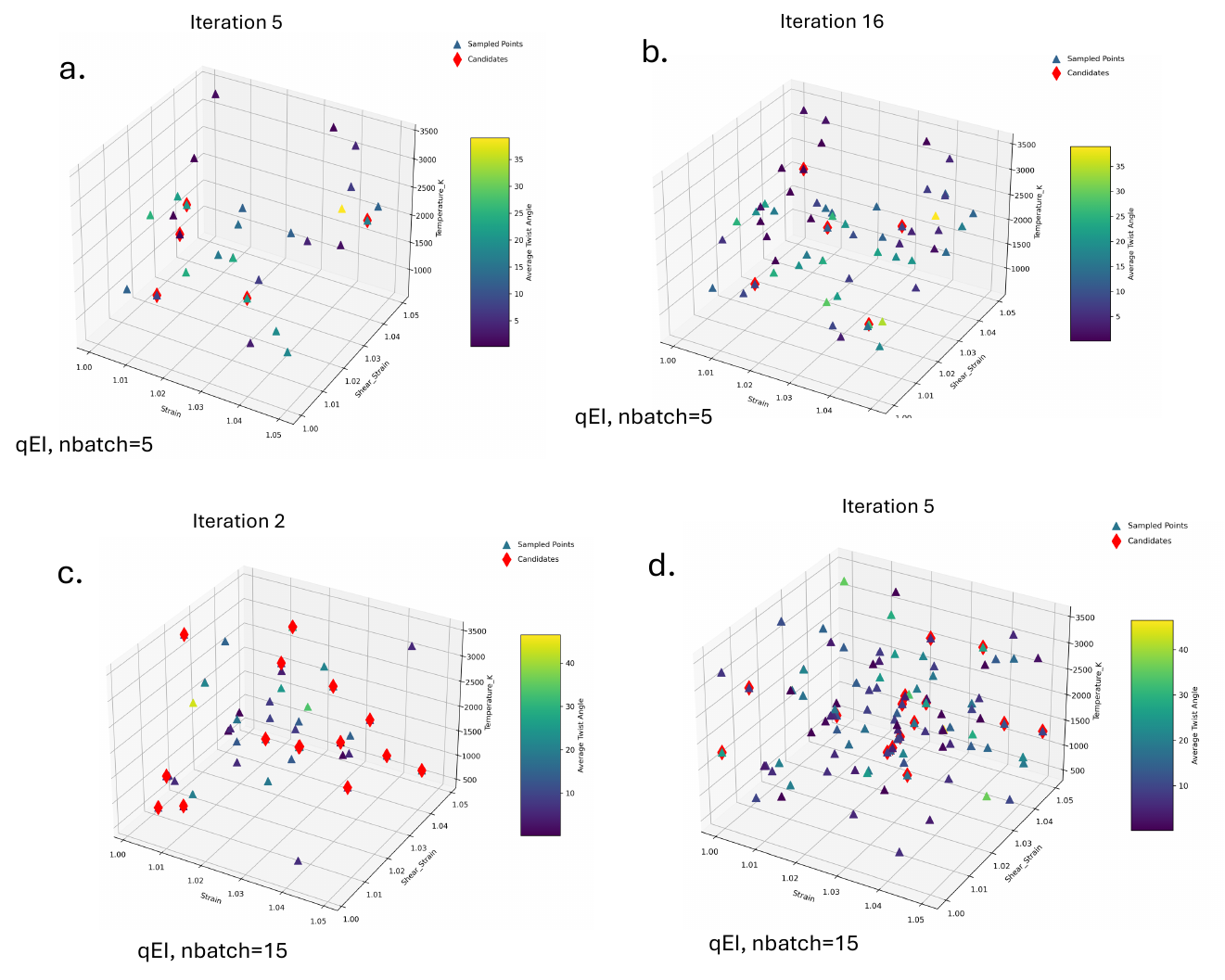}
\caption{\textbf{Evolution of batch sampling for targeted recrystallization} Snapshots showing how the parameter space $\{T_i, \epsilon_{lat \ i}, \epsilon_{sh \ i}\}$ is sampled during the BO for $n_{batch}=5$ in \textbf{(a,b)} and $n_{batch}=15$ in \textbf{(c,d).}}
\label{fig:SI-sampling_evol}
\end{figure}

\begin{figure}[hbt!]
\centering
\includegraphics[height=.45\linewidth,width=.75\linewidth,angle=0]{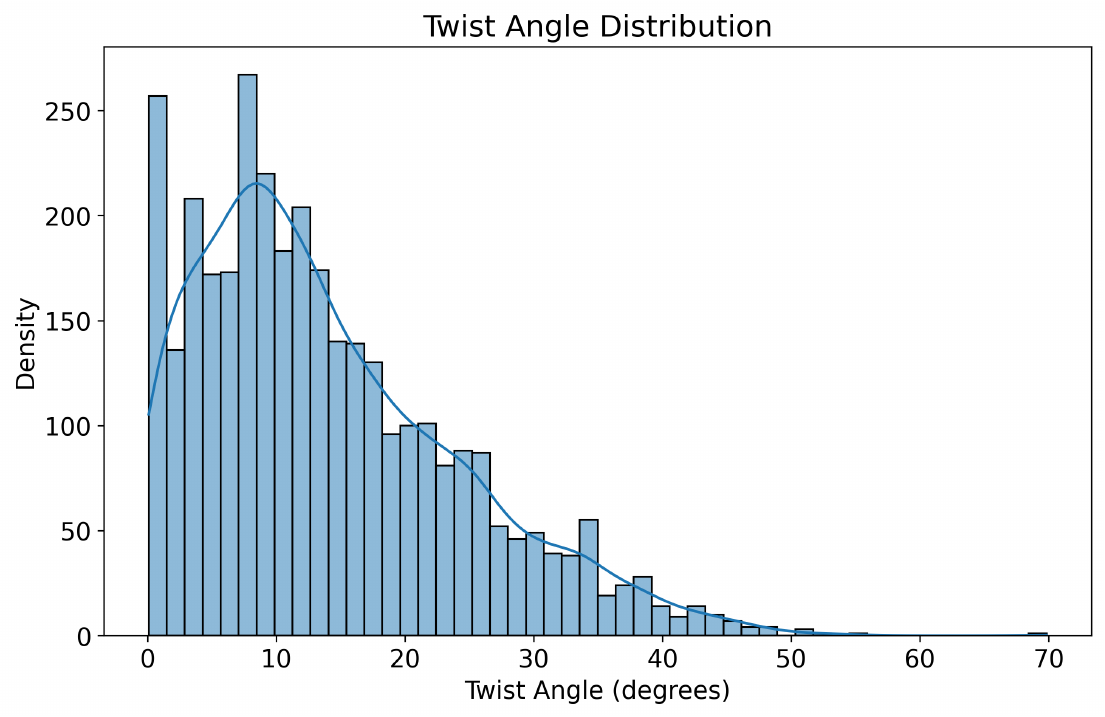}
\caption{\textbf{Histogram of twist angles sampled in a grid exploratory
fashion.} It is evident from the distribution of over 3000 total hetero-structures shown in the
figure that setting target twist angles beyond 10◦ are increasingly nontrivial to find optimal
growth parameters for our "on-demand" sampling results for angles at the tail-end e.g., 25 deg. .}
\label{fig:SI-hist}
\end{figure}

\begin{figure}[hbt!]
\centering
\includegraphics[height=1.2\linewidth,width=\linewidth,angle=0]{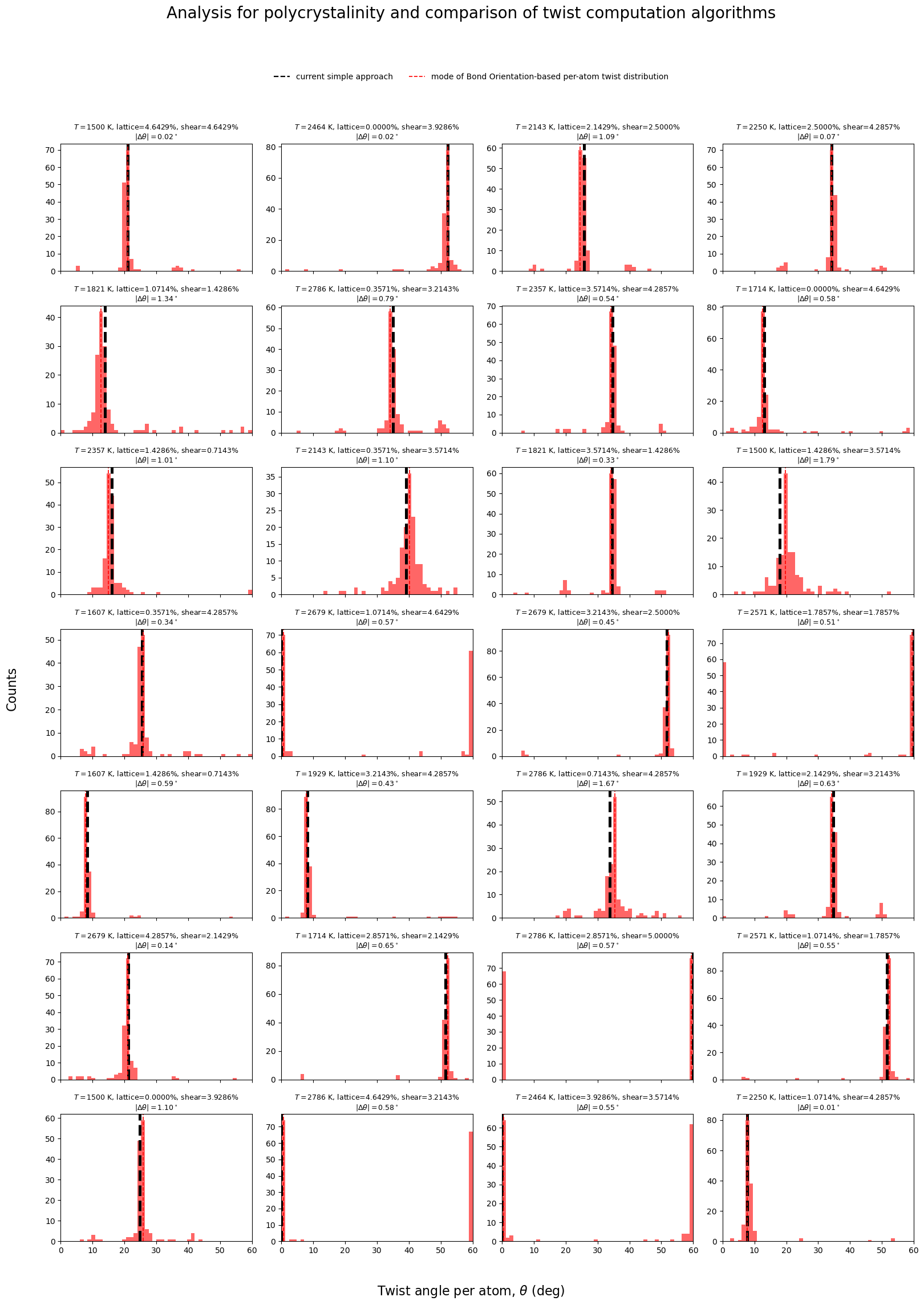}
\caption{\added{\textbf{Evidence of single crystallinity and validation of current twist computation
approach (black dash) with bond-orientation based per-atom calculations (red histograms)}}}
\label{fig:single-crystalline}
\end{figure}

\end{document}